\documentclass[10pt,letterpaper]{article}
\usepackage{opex3}
\usepackage{amsmath} 
\usepackage{color}
\usepackage[percent]{overpic}
\usepackage{cite}
\usepackage{upgreek}
\usepackage{fancyhdr}
\pdfoutput=1

\newcommand*{\mos}{MoS$_2$ }
\newcommand*{\mosN}{MoS$_2$}
\newcommand*{\mose}{MoSe$_2$ }
\newcommand*{\moseN}{MoSe$_2$}
\newcommand*{\ws}{WS$_2$ }
\newcommand*{\wsN}{WS$_2$}

\newcommand*{\wseN}{WSe$_2$}
\newcommand*{\apx}{$\sim$}
\newcommand{\enquote}[1]{``#1''}
\begin{document}
\title{Wideband saturable absorption in few-layer molybdenum diselenide (MoSe$_2$) for Q-switching Yb-, Er- and Tm-doped fiber lasers}
\author{R. I. Woodward,$^{1,\dagger,*}$ R. C. T. Howe,$^{2,\dagger}$ T.~H.~Runcorn,$^1$ G.~Hu,$^2$ F.~Torrisi,$^2$ E.~J.~R.~Kelleher,$^1$ and T. Hasan$^{2}$}
\address{$^1$Femtosecond Optics Group, Department of Physics, Imperial College London, SW7 2AZ, UK\\
$^2$Cambridge Graphene Centre, University of Cambridge, Cambridge, CB3 0FA, UK\\
$^\dagger$ These authors contributed equally to this work.
}
\email{$^*$r.woodward12@imperial.ac.uk} 
\begin{abstract}
We fabricate a free-standing molybdenum diselenide (\moseN) saturable absorber by embedding liquid-phase exfoliated few-layer \mose flakes into a polymer film.
The \moseN-polymer composite is used to \mbox{Q-switch} fiber lasers based on ytterbium (Yb), erbium (Er) and thulium (Tm) gain fiber, producing trains of microsecond-duration pulses with kilohertz repetition rates at 1060~nm, 1566~nm and 1924~nm, respectively.
Such operating wavelengths correspond to sub-bandgap saturable absorption in \moseN, which is explained in the context of edge-states, building upon studies of other semiconducting transition metal dichalcogenide (TMD)-based saturable absorbers.
Our work adds few-layer \mose to the growing catalog of TMDs with remarkable optical properties, which offer new opportunities for photonic devices.
\end{abstract}


\section{Introduction}
Transition metal dichalcogenides (TMDs) are a family of more than 40 layered materials with general formula MX$_2$, where M is a transition metal, and X is a chalcogen (i.e. a group VI element such as sulfur or selenium)~\cite{Wilson1969a}. 
Their structure consists of quasi-2d layers weakly bound together by van der Waals forces. 
In each layer, a plane of transition metal atoms is covalently bonded between two planes of chalcogen atoms. 
TMDs have a diverse range of properties and include metallic (e.g. NbS$_2$), semiconducting (e.g. \mosN) and insulating (e.g.~HfS$_2$) materials~\cite{Wilson1969a}. 

Despite many fundamental studies of the structure and properties of TMDs in the 1960s and 70s~\cite{Wilson1969a, James1963, Evans1971, Doviak1972, Joensen1986, Roxlo1987}, including reports of monolayer~\cite{Joensen1986} and few-layer crystals~\cite{Evans1971}, the lack of suitable processing and characterization techniques for such materials meant that their unique optical characteristics were not exploited for practical technologies.
At present, semiconducting TMDs (\mosN, \moseN, \wsN, \wseN, MoTe$_2$ etc.) are experiencing renewed interest due to their remarkable layer-dependent optoelectronic properties, which hold great promise for future photonic devices\cite{Wang2012_nn}.
These include a transition from an indirect to direct bandgap at visible / near-infrared wavelengths when moving from bulk to monolayer form, in addition to strong photoluminescence, high nonlinearity and ultrafast carrier dynamics for mono- and few-layer forms~\cite{Wang2012_nn}. 
The direct bandgaps of monolayer semiconducting TMDs in the visible and near-infrared offer advantages over graphene, a zero-gap material, for many optoelectronic applications. 
Additionally, the layer- and size-dependent characteristics of these TMDs allow great flexibility for engineering their optical properties~\cite{Wang2012_nn}. 

Molybdenum disulfide (\mosN) is perhaps the most widely studied TMD to date~\cite{Joensen1986, Roxlo1987, Wang2012_nn, Mak2010_prl}.
Recent work has demonstrated wideband saturable absorption in few-layer \mosN~\cite{Wang2013, Wang2014b, Zhou2014, Ouyang2014}, suggesting that semiconducting TMDs could be a promising class of saturable absorber (SA) for short laser pulse generation~\cite{Woodward_prj_2015}. 
Numerous studies have exploited this SA response, demonstrating mode-locked and Q-switched lasers (with both bulk and fiber gain media) from 1030 to 2100 nm~\cite{Woodward_prj_2015,Wang2014_am,Zhang2014, Woodward_cleo14_mos2, Woodward_oe_2014_mos2, Khazaeinezhad2015, Luo2014, Liu2014b,Xia2014a,Xu2014d,Zhang_nr_2014, Huang2014} either at discrete wavelengths or with broadband tunability~\cite{Zhang2014,Woodward_oe_2014_mos2,Huang2014}.
It has been proposed that the ability of a single \mos SA device to operate at numerous wavelengths, even in spectral regions corresponding to sub-bandgap photon energies, is due to edge-state absorption in the material~\cite{Woodward_oe_2014_mos2,Woodward_prj_2015}.

More recently, other \emph{sulfide}-based TMDs such as tungsten disulfide (\wsN) have emerged as candidate materials for future saturable absorbers: \ws has recently been demonstrated to mode-lock and Q-switch fiber lasers~\cite{Wu2014_arx, Mao2015, Yan2015,Kassani2015}.
However, \emph{selenide}-based TMDs (\moseN, \wseN, etc.) have yet to be fully explored and could offer advantages over sulfide TMDs for certain applications, for instance where a narrower gap semiconductor is required (heavier chalcogenide atoms lead to reduced bandgap energies~\cite{Wang2012_nn,Zhou2015}). 
Therefore, further investigation to understand the optical properties of few-layer transition metal selenides is required before they can be exploited in practical systems. 

Few-layer molybdenum diselenide (\moseN) flakes have been fabricated by mechanical exfoliation~\cite{Tongay2012}, ultrasound-assisted liquid-phase exfoliation (UALPE)~\cite{Coleman2011a} and chemical vapor deposition (CVD)~\cite{Shim2014, Shaw2014, Wang2015_acs} techniques.
Bulk \mose has a \apx1.1~eV ($\sim$1128~nm) indirect bandgap, which increases to a \apx1.41~eV ($\sim$879~nm) indirect gap for 8-layer \moseN~\cite{Zhang2014c} and a direct \apx1.55~eV ($\sim$800~nm) gap for monolayer \moseN~\cite{Tongay2012,Zhang2014c}.
The nonlinear optical properties of few-layer \mose have been studied by Wang et al. for UALPE \mose dispersed in cyclohexylpyrrolidone (CHP)~\cite{Wang2014b}, and by Dong et al. in N-methyl-pyrrolidone (NMP)~\cite{Dong2015}; both saturable absorption and optical limiting behavior were reported. 
During the review process, mode-locking of a fiber laser using few-layer \mose was reported \cite{Luo2015a}, although only for operation at a single wavelength of 1558~nm.

Here, we produce a few-layer \moseN-polyvinyl alcohol (PVA) composite SA for short-pulse generation. 
Fiber lasers based on ytterbium- (Yb), erbium- (Er) and thulium-doped (Tm) gain media are demonstrated, operating at 1060~nm, 1566~nm and ~1924~nm, respectively, which generate stable trains of Q-switched pulses when the \moseN-PVA SA is integrated into the cavity.

\section{Few-layer \mose fabrication and integration into a saturable absorber device}
\begin{figure}[bp]
\centerline{\includegraphics[width=\columnwidth]{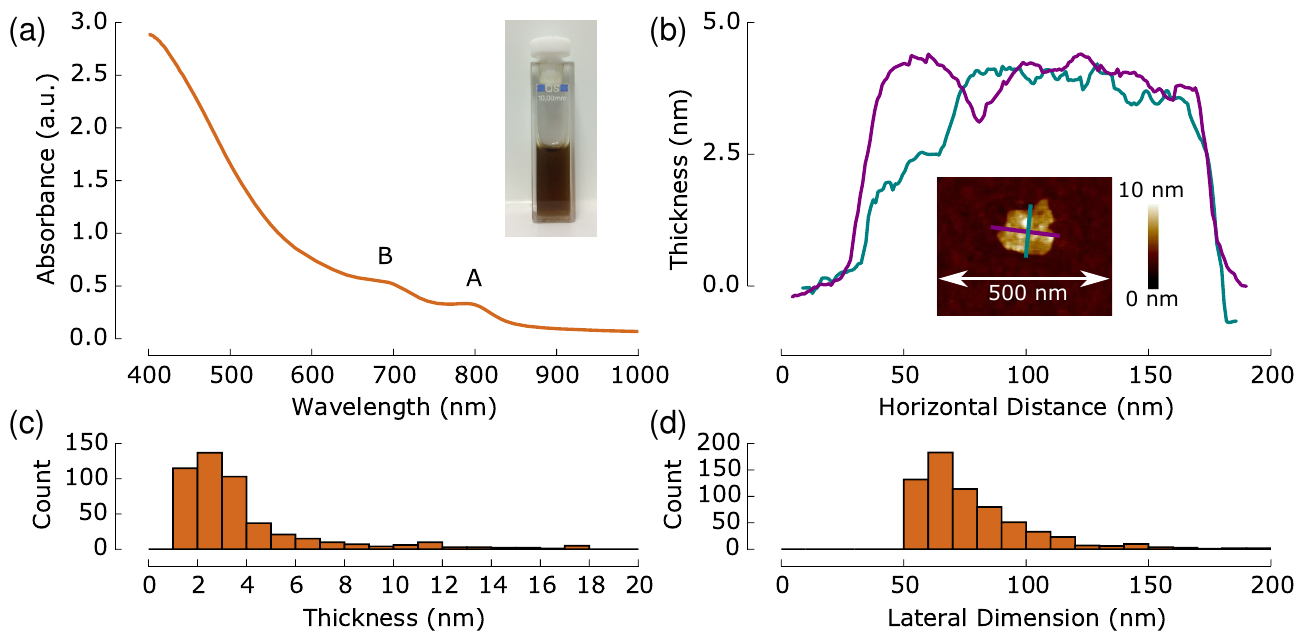}}
	\caption{(a) Linear optical absorption of \mose flakes in a 10\% (v/v) dispersion, showing characteristic excitonic peaks (A \& B) (inset: photograph of the diluted dispersion). AFM characterization of \mose flakes: (b) AFM image of a typical flake and two thickness profiles; (c) distribution of flake thicknesses; (d) distribution of lateral flake dimensions.}
	\label{fig:dispersion}
\end{figure}

We use an ultrasound-assisted liquid phase exfoliation (UALPE) technique to produce few-layer \mose nanoflakes from their bulk crystals. 
UALPE exploits cavitation arising from high frequency pressure variation.
These microbubbles subsequently collapse due to instability, resulting in high shear forces~\cite{Mason2002book}.
This exfoliates flakes from bulk crystals by overcoming the interlayer van der Waals forces. 

In general, solvents matching the experimentally-determined Hansen solubility parameters~\cite{Hansen2007} of layered materials support their exfoliation and stabilization~\cite{Coleman2012}. 
This offers a more refined understanding than considering only the surface energies or Hildebrand solubility parameters of the solvents and materials~\cite{Coleman2012,Cunningham2012}. 
Experimental evidence suggests that NMP and CHP are two of the more suitable solvents for TMDs~\cite{Cunningham2012}.
However, it is challenging to process these high boiling point ($>$150$^\circ$C) solvents for the fabrication of SA composites. 
Low boiling point solvents, e.g. water, are thus more desirable to exfoliate \moseN, although large mismatch of Hansen solubility parameters of water~\cite{Hansen2007} with \mose (and other layered materials)~\cite{Coleman2012} dictates the use of surfactants for its exfoliation and stabilization in water. 

For layered materials, the molecular structure of surfactants plays an important role. 
The most effective are quasi-2d surfactants such as bile salts (e.g. sodium cholate) due to their planar structure with a hydrophobic and a hydrophilic side. 
The hydrophobic side of the bile salt surfactant molecules adsorbs onto the flat surface of the layered materials, while the hydrophilic side creates an effective surface charge around the flake~\cite{Hasan2010}. 
The exfoliated layered materials are thus stabilized against re-aggregation by the Coulomb repulsion resulting from this surface charge.
Among the different bile salt surfactants, we choose sodium deoxycholate (SDC) due to its high hydrophobic index\cite{Hasan2010}. 
We prepare the \mose dispersion by sonicating 100 mg of \mose crystals (Sigma Aldrich, 325 mesh) with 70~mg SDC (Sigma Aldrich, m.w. $\sim$1200-5000) in 10~mL deionized water for 12 hours at 15$^\circ$C. 
The resultant dispersion is centrifuged at 4000~rpm ($\sim$1500\emph{g}) for 1~hour to sediment the unexfoliated large flakes. 
The upper 70\% of the centrifuged dispersion, enriched with few layer \mose flakes, is decanted for characterization and composite fabrication.

The optical absorption spectrum of the \mose dispersion, diluted to 10\% (v/v) to avoid scattering losses, shows two absorption peaks labeled as A ($\sim$800 nm) and B ($\sim$710~nm) [Fig.~\ref{fig:dispersion}(a)].
These peaks correspond to the excitons from the two spin-orbit split transitions at the \emph{K} point of the Brillouin zone\cite{Beal1972,Li2014i}. 
To estimate the concentration of the dispersed \moseN, we use the Beer-Lambert law, A$_{\lambda}=\alpha_{\lambda}lc$, where A$_{\lambda}$ is the absorbance at wavelength $\lambda$, $\alpha_{\lambda}$ is the absorption coefficient at $\lambda$, \emph{c} is the concentration, and $l$ is the optical path of incident light through the dispersion. 
First, we experimentally estimate $\alpha_{\lambda}$ for \mose at $\lambda=$710~nm following the procedure developed in Ref.~\cite{Zhang_nr_2014} for \mosN. 
This gives $\alpha_{710nm}=615$~Lg$^{-1}$m$^{-1}$, from which we estimate the concentration of the undiluted \mose as 0.78 gL$^{-1}$.
\begin{figure}[bp]
\centerline{\includegraphics[width=\columnwidth]{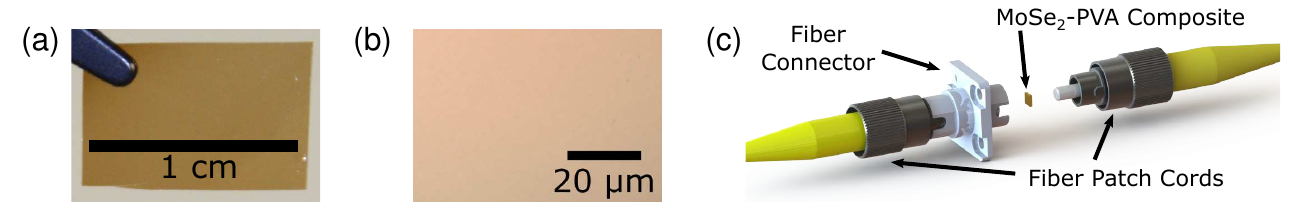}}
	\caption{(a) Photograph of \moseN-PVA composite. (b) Optical micrograph showing absence of aggregates. (c) Schematic showing integration of \mose SA device between two fiber patch cords.}
	\label{fig:composite}
\end{figure}

\begin{figure}[tbp]
\centerline{\includegraphics[width=\columnwidth]{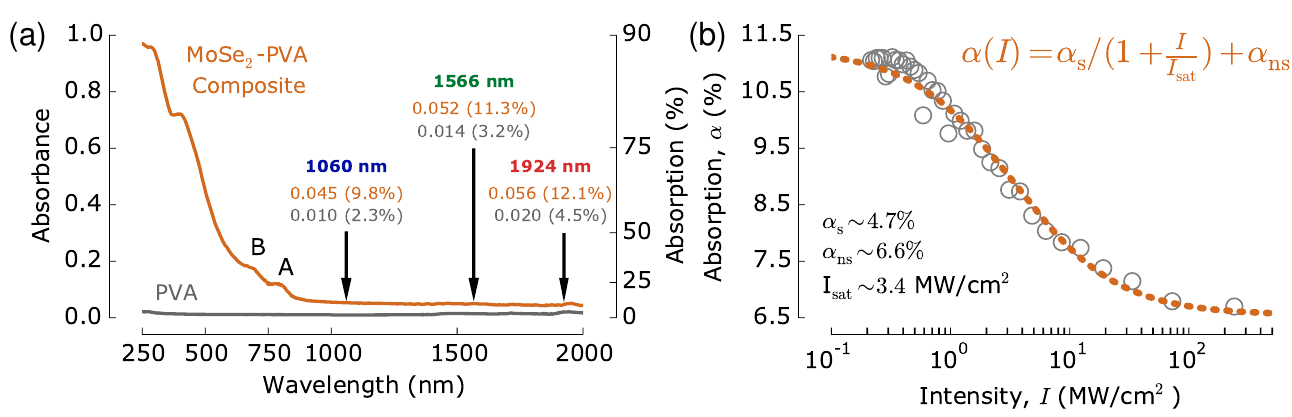}}
	\caption{Few-layer \moseN-PVA SA characterization: (a) linear optical absorption, with highlighted absorption values at the operating wavelengths of our \moseN-based Q-switched lasers; (b) nonlinear optical absorption (at 1566~nm) measured by a Z-scan technique and well fitted by a two-level SA model~\cite{Garmire2000},where $\alpha$ is absorption, $I$ is incident intensity, $\alpha_\text{s}$ is modulation depth, $\alpha_\text{ns}$ is nonsaturable absorption and $I_\text{sat}$ is saturation intensity.}
	\label{fig:sa_properties}
\end{figure}

The thickness distribution of exfoliated \mose flake dimensions is measured via atomic force microscopy (AFM), shown in Figs.~\ref{fig:dispersion}(b)--\ref{fig:dispersion}(d). 
Samples for AFM are prepared by diluting the \mose dispersion to 5\% (v/v), drop-casting on to a Si/SiO$_2$ wafer and then rinsing with deionized water to remove the surfactant. 
The average flake thickness is 3.8~$\pm$~0.1~nm, while 80\% of the flakes have thickness $\leq$5~nm [Fig.~\ref{fig:dispersion}(c)], corresponding to $\leq$6--7 layers for the majority of the flakes, assuming 0.9--1.0 nm measured thickness for a monolayer flake~\cite{Tongay2012}, and 0.65--0.7 nm increase for each additional layer \cite{Zhang2014c}. 
The average lateral dimension of the exfoliated flakes is 80~$\pm$~5~nm, resulting in a high edge-to-surface area ratio [Fig.~\ref{fig:dispersion}(d)].

The free-standing \moseN-polymer composite SA is fabricated by following a procedure commonly adopted for other layered material-based polymer composite SAs such as graphene\cite{Hasan2010} and \mos\cite{Zhang_nr_2014, Woodward_oe_2014_mos2}.
Briefly, 1.5~mL of the \mose dispersion is homogeneously mixed with 2 mL 5 wt\% aqueous PVA solution.
The mixture is poured into a petri dish and dried at room temperature in a desiccator. 
Slow evaporation of water over several days forms a $\sim$30~$\upmu$m thick free-standing composite SA film [Fig.~\ref{fig:composite}(a)].
An optical microscope image of the composite verifies the absence of aggregates and voids [Fig.~\ref{fig:composite}(b)], which could otherwise increase the non-saturable scattering loss.
For comparison, a \apx30~$\upmu$m thick pure PVA film is fabricated using a similar method, but without the \mose flakes.

The linear optical absorption profile of the \moseN-PVA film shows the characteristic \mose excitonic peaks (A \& B) and a non-negligible absorbance throughout the near-infrared ($>$800~nm), as highlighted in Fig.~\ref{fig:sa_properties}(a).
The composite film absorption is $\sim$9.8\%, $\sim$11.3\% and $\sim$12.1\% at 1060~nm, 1566~nm and 1924~nm, respectively (the three operating wavelengths of the lasers in Section 3), demonstrating a marked increase over the absorption values for the pure PVA film of $\sim$2.3\%, $\sim$3.2\% and $\sim$4.5\%.
The nonlinear optical response of the few-layer \moseN-PVA composite is measured using an open-aperture Z-scan technique~\cite{Sheik-Bahae1990a}: the composite is swept through the focus of a beam of ultrashort pulses from a mode-locked fiber laser (750~fs pulse duration, 1566~nm wavelength, 7.5~MHz repetition rate) and the transmitted power is recorded as a function of incident intensity on the device (in addition to a reference power signal for normalization). Our few-layer \mose sample exhibits a strong saturable absorption response [Fig.~\ref{fig:sa_properties}(b)], well-fitted by the two-level SA model~\cite{Garmire2000}.
From this fit, the following SA parameters can be extracted: saturation intensity, $I_\text{sat}$ \apx 3.4 MW/cm$^2$; modulation depth, $\alpha_\text{s}$ \apx 4.7\% and nonsaturable absorption, $\alpha_\text{ns}$ \apx 6.6\%.
Such SA parameters are similar in magnitude to previously reported saturable absorbers based on few-layer \mosN~\cite{Zhang_nr_2014,Woodward_prj_2015}.
We note that the Z-scan pulse source at 1566~nm corresponds to a photon energy of 0.79~eV, less than the energy required for single photon excitation of carriers across the bandgap of few-layer \moseN~\cite{Tongay2012}.
A proposal for the mechanism of the observed sub-bandgap saturable absorption is discussed in Section~\ref{sec:discuss}.

\section{Q-switched fiber lasers using few-layer \mose SA}
The demonstrated sub-bandgap saturable absorption of our few-layer \moseN-PVA composite at 1566~nm indicates that the device could Q-switch an Er-doped fiber laser cavity to generate short optical pulses, which are needed for many applications such as fundamental research and industrial materials processing and micromachining.
The \moseN-PVA SA is integrated into an Er-doped fiber laser, in addition to Yb- and Tm-doped fiber lasers to explore the potential of exploiting sub-bandgap saturable absorption for short-pulse generation at different near-infrared wavelengths, using a single \mose SA. 
For each laser cavity, a ring design is adopted, including a polarization-independent isolator, output coupler and polarization controller, in addition to the fiber amplifier [Figs.~\ref{fig:yb_laser}(a), \ref{fig:er_laser}(a) and \ref{fig:tm_laser}(a)].
The Yb and Er amplifiers consist of double-clad Yb and Er fiber, which are 1.5~m and 11.6~m long, respectively, pumped by multimode 965~nm laser diodes.
The Tm fiber amplifier is formed of a 2.9~m length of single-clad Tm-doped fiber, pumped by a 1550~nm continuous wave fiber laser.
The total lengths of the Yb, Er, and Tm fiber-based cavities are $\sim$15.9~m, $\sim$27.4~m and $\sim$13.4~m, respectively.
A 1 mm~$\times$~1~ mm piece of the \mose SA film is integrated into the cavity by placing it between two fiber patch cords [Fig.~\ref{fig:composite}(c)].

\subsection{Q-switched Yb:fiber laser characterization} 
\begin{figure}[htbp]
\centerline{\includegraphics[width=\columnwidth]{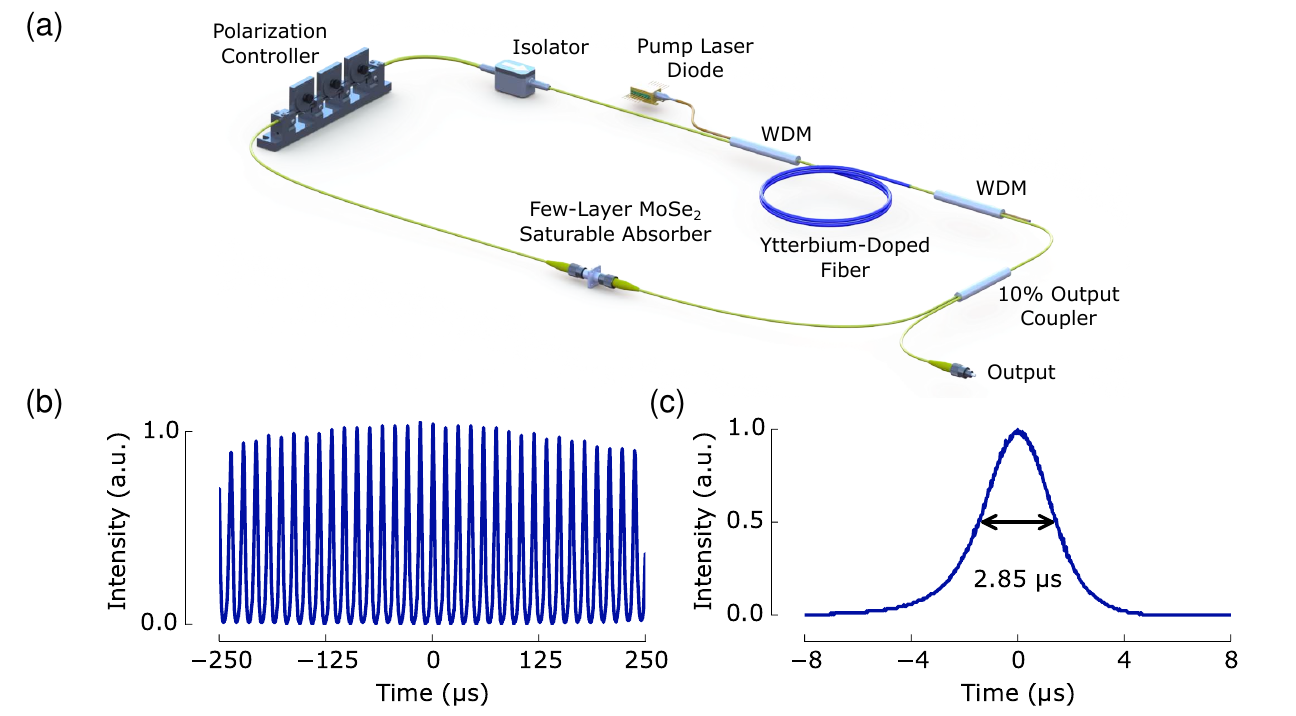}}
	\caption{Yb:fiber Q-switched laser: (a) cavity schematic; (b) output pulse train at 65.0~kHz repetition rate; (c) pulse profile with 2.85~$\upmu$s duration for 8.0 mW average output power.}
	\label{fig:yb_laser}
\end{figure}

Self-starting Q-switching is observed from the Yb-doped fiber laser, generating a steady train of pulses [typical pulse characteristics shown in Figs.~\ref{fig:yb_laser}(b) and \ref{fig:yb_laser}(c)], centered at 1060~nm [Fig.~\ref{fig:spectra}(a)].
By changing the pump power, the average output power can be varied from 6.26~mW (corresponding to the threshold for Q-switched emission; below this, the laser output is a continuous wave) to 8.72~mW. 
The steady-state output pulse properties depend on the cavity gain and absorber saturation dynamics, coupling the pump power to the pulse duration and repetition rate. 
With increasing power, the pulse repetition frequency is increased from 60.0~kHz to 74.9~kHz and the duration decreased from 4.6~$\upmu$s to 2.8~$\upmu$s [Fig.~\ref{fig:reprate_duration}(a)].
At the maximum average output power, the pulse energy is 116~nJ.

\subsection{Q-switched Er:fiber laser characterization} 
\begin{figure}[htbp]
\centerline{\includegraphics[width=\columnwidth]{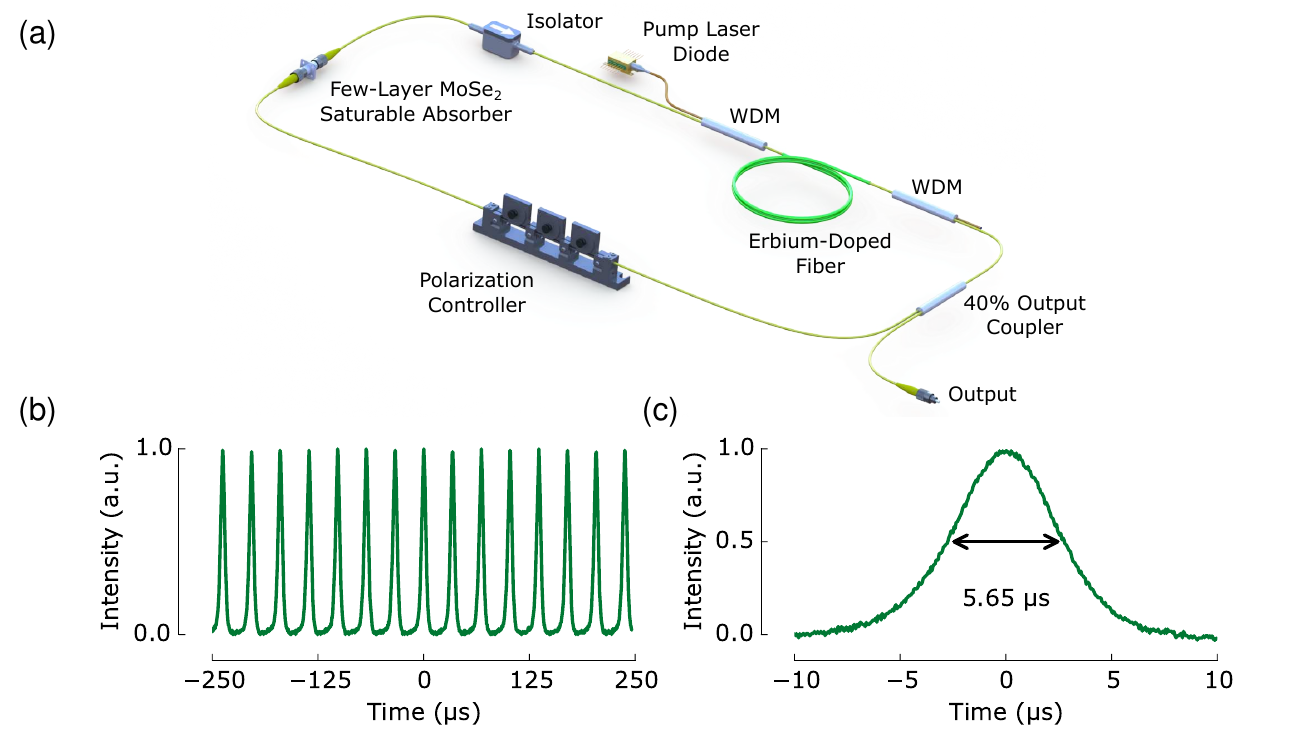}}
	\caption{Er:fiber Q-switched laser: (a) cavity schematic; (b) output pulse train at 29.4~kHz repetition rate; (c) pulse profile with 5.65~$\upmu$s duration for 24.5 mW average output power.}
	\label{fig:er_laser}
\end{figure}

The Er-doped fiber laser incorporating the few-layer \moseN-PVA composite exhibits self-starting Q-switching at an average output power of 18.9~mW, operating at 1566~nm [Fig.~\ref{fig:spectra}(b)]. 
Typical output pulse train properties are shown in Figs.~\ref{fig:er_laser}(b) and \ref{fig:er_laser}(c).
By varying the pump power, the repetition rate could be changed from 26.5~kHz to 35.4~kHz and the duration from 7.9~$\upmu$s to 4.8~$\upmu$s as the average output power increased from 18.9~mW to 29.2~mW [Fig.~\ref{fig:reprate_duration}(b)]. 
At the highest output power, the pulse energy is 825~nJ.

\subsection{Q-switched Tm:fiber laser characterization} 
\begin{figure}[htbp]
\centerline{\includegraphics[width=\columnwidth]{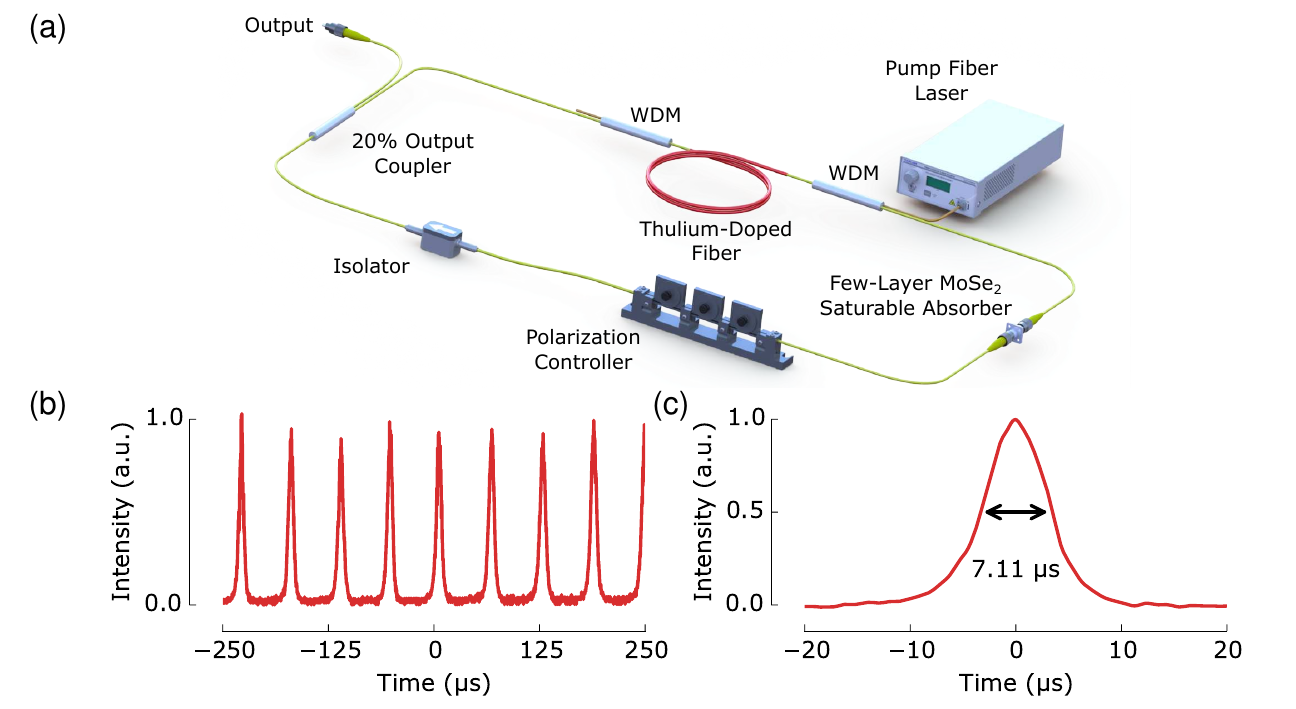}}
	\caption{Tm:fiber Q-switched laser: (a) cavity schematic; (b) output pulse train at 16.9~kHz repetition rate; (c) pulse profile with 7.11~$\upmu$s duration for 0.79~mW average output power.}
	\label{fig:tm_laser}
\end{figure}

\begin{figure}[tbp]
\centerline{\includegraphics[width=\columnwidth]{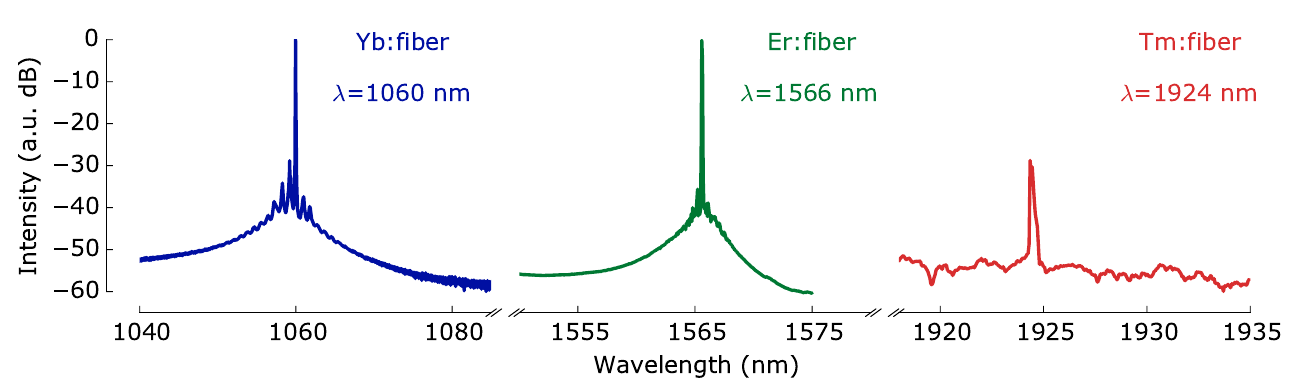}}
	\caption{Spectra for Q-switched operation of few-layer \mose Q-switched Yb- and Er- and Tm-doped fiber lasers; the Yb and Er spectrum were measured using an optical spectrum analyzer (0.02~nm resolution), while the Tm spectrum was measured on a spectrometer ($\sim$0.1~nm resolution).}
	\label{fig:spectra}
\end{figure}

\begin{figure}[tbp]
\centerline{\includegraphics[width=\columnwidth]{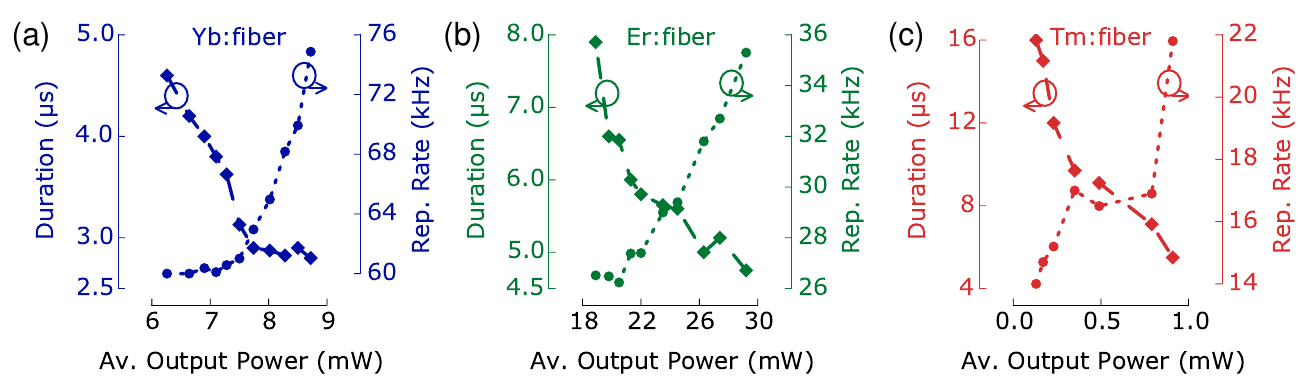}}
	\caption{Variation of pulse duration and repetition rate with average output power for (a) Yb-, (b) Er-, and (c) Tm-doped fiber lasers, Q-switched with a few-layer \mose SA.}
	\label{fig:reprate_duration}
\end{figure}

Continuous-wave lasing at 1924~nm is initially observed as the pump power of the Tm-doped fiber laser is increased, until the Q-switching threshold is reached, corresponding to 0.13~mW average output power [Fig.~\ref{fig:spectra}(c)].
Beyond threshold, a steady train of Q-switched pulses is generated [Figs.~\ref{fig:tm_laser}(b) and ~\ref{fig:tm_laser}(c)].
As the pump power is increased, the average output power increases and the Q-switched pulse duration reduces from 16.0~$\upmu$s to 5.5~$\upmu$s while the repetition rate increases from 14.0~kHz to 21.8~kHz [Fig.~\ref{fig:reprate_duration}(c)].
The maximum pulse energy is 42~nJ.

\subsection{Discussion}
\label{sec:discuss}
The few-layer \moseN-PVA composite enables Q-switched operation in fiber laser cavities at 1060~nm, 1566~nm and 1924~nm.
The lasers were also operated with \apx30~$\upmu$m-thick pure PVA film (fabricated in a similar way to the \moseN-PVA composite, but without \mose flakes) sandwiched between the fiber patch cords, although in this case Q-switching was not observed at any power level or polarization controller position, confirming that the saturable absorption arises from the few-layer \mose material.

While bulk \mose has an indirect bandgap of \apx1.1~eV (\apx1128~nm)~\cite{Tongay2012,Zhang2014c}, a recent angle-resolved photoemission spectroscopy (ARPES) study on molecular beam epitaxy \mbox{(MBE)-grown} \mose samples demonstrated a transition from an indirect bandgap of \apx1.41~eV (\apx880~nm) for 8-layer \mose to a direct bandgap of \apx1.58~eV (\apx784~nm) for monolayer \mose~\cite{Zhang2014c}.

AFM statistics, considering $>$400 flakes of our UALPE \mose samples, show that 80\% of the flakes consists of $\leq$6--7 layers. 
Based on the above experimental data and the ARPES study by Ref. \cite{Zhang2014c}, we estimate an indirect bandgap just above $\sim$1.41 eV for these flakes.
Therefore, single-photon optical absorption from the few-layer \mose samples above 880~nm is not expected.
However, our linear absorption measurement indicates \apx11\% absorption at these near-infrared wavelengths ($>$880~nm), of which $>$70\% (\apx8\% absolute absorption) is due to the addition of \mose flakes, compared to the pure PVA film [Fig.~\ref{fig:composite}(a)].
While this measurement could have also been affected by scattering loss, the sub-bandgap absorption observed in our UALPE \mose is verified by the Z-scan measurement at 1566~nm and demonstration of few-layer \moseN-based Q-switched lasers at 1060~nm (1.16~eV), 1566~nm (0.79~eV) and 1924~nm (0.64~eV).
Indeed, similar observations of sub-bandgap saturable absorption have been reported for few-layer \mosN~\cite{Wang2013, Wang2014b, Zhou2014, Ouyang2014, Woodward_prj_2015, Wang2014_am,Zhang2014,Woodward_cleo14_mos2, Woodward_oe_2014_mos2, Khazaeinezhad2015, Luo2014, Liu2014b,Xia2014a, Xu2014d, Zhang_nr_2014} and \ws~\cite{Wu2014_arx, Mao2015, Yan2015,Kassani2015}.

We recently proposed an explanation for this phenomenon based on edge-states (in the context of \mos SA devices)~\cite{Woodward_oe_2014_mos2,Zhang_nr_2014,Woodward_prj_2015}. 
This mechanism is supported by early photothermal deflection spectroscopy studies of few-layer \mos samples ~\cite{Roxlo1987}.
By comparing \mos samples with different microcrystal sizes and lithographically textured single crystals, Ref.~\cite{Roxlo1987} demonstrated increased sub-bandgap absorption in samples with a greater amount of crystal edges. 
This suggests that states at the edges of the \mos flakes play a significant role in defining the material band structure and enable absorption by exciting electrons to the edge-state levels within the material bandgap. 
This edge-induced sub-bandgap absorption can be saturated by Pauli blocking at high incident intensities, which enables the material to act as a Q-switch in a laser cavity~\cite{Woodward_oe_2014_mos2,Zhang_nr_2014,Woodward_prj_2015}.
We note that Wang et al. have also proposed a complementary explanation for this phenomenon, supported by theoretical bandgap studies, based on crystallographic defect states~\cite{Wang2014_am}.

Generalizing the above discussion to other semiconducting few-layer TMD materials, we propose that a high edge-to-surface area ratio, can lead to sub-bandgap absorption in these materials. 
While this is not fully understood at present, crystallographic defects, dislocations, grain boundaries $\emph{etc}$ in semiconducting TMD samples may also contribute to sub-bandgap absorption. 
This could also explain observations of saturable absorption in CVD-grown \mos samples~\cite{Xia2014a}, which, in general, are expected to have less edge-to-surface area ratio than those prepared by UALPE.
Our UALPE \mose flakes with average lateral dimensions of 80~$\pm$~5~nm have a high edge-to-surface area ratio, which we propose to be the primary origin of the optical absorption below the material bandgap.

Additionally, Kumar et al. recently reported absorption saturation in both bulk and monolayer \mose as part of a femtosecond transient absorption study, which they attributed to exciton-exciton interactions under intense pumping~\cite{Kumar2014}. 
However, in this case the pump wavelength was 750~nm, tuned to the excitonic resonance.
Therefore, while this excitonic mechanism may contribute to saturable absorption at wavelengths corresponding to the bandgap, all three of our lasers operate at wavelengths with insufficient photon energies for resonant exciton excitation.
Consequently, we believe that excitonic effects are not involved in the Q-switching of our lasers.
Instead, we attribute the few-layer \mose intensity-dependent absorption to saturable edge states.

While the direct visible bandgap of monolayer and few-layer \mose suggests suitability for optical applications in the visible spectral region, the edge-state driven wideband absorption in the infrared highlights that few-layer \mose could be a very versatile material for developing future laser-based photonic technologies.

\section{Conclusion}
In summary, we have reported the production of few-layer \mose flakes by ultrasound-assisted liquid-phase exfoliation and the fabrication of a saturable absorber device by embedding these \mose flakes into PVA.
For the first time, we have demonstrated that a selenide-based transition metal dichalcogenide is able to generate short laser pulses by Q-switching: we developed ytterbium-, erbium- and thulium-doped fiber lasers, Q-switched by few-layer \moseN, operating at 1060~nm, 1566~nm and 1924~nm, respectively.
The wideband saturable absorption was explained in the context of sub-bandgap edge states, extending our existing proposed mechanism for this phenomenon to the wider class of semiconducting transition metal dichalcogenides.
This adds few-layer \mose to the growing library of nanomaterials with exceptional optical properties that offer exciting opportunities for future photonic applications.

\section*{Acknowledgments}
We thank J. R. Taylor and S. V. Popov for fruitful discussions.
EJRK and TH acknowledge support from the Royal Academy of Engineering (RAEng), through RAEng Fellowships.

\end{document}